\documentclass[12pt,thmsa]{article}
\usepackage{amsfonts}

\usepackage{sw20aip}

\input{tcilatex}
\begin{document}

\author{Ezra T. Newman \\
%EndAName
Dept of Physics and Astronomy, \\
Univ.of Pittsburgh, Pgh., PA 15260\\
newman@pitt.edu}
\title{Maxwell Fields \& Shear-Free Null Geodesic Congruences}
\date{Feb.12, 2004}
\maketitle

\begin{abstract}
We study and report on the class of vacuum Maxwell fields in Minkowski space
that possess a non-degenerate, diverging, principle null vector field (null
eigenvector field of the Maxwell tensor) that is tangent to a shear-free
null geodesics congruence. These congruences can be either surface forming
(the tangent vectors being proportional to gradients) or not, i.e., the
twisting congruences. In the non-twisting case, the associated Maxwell
fields are precisely the Lienard-Wiechert fields, i.e., those Maxwell fields
arising from an electric monopole moving on an arbitrary worldline. The null
geodesic congruence is given by the generators of the light-cones with apex
on the world-line. The twisting case is much richer, more interesting and
far more complicated.

\quad In a twisting subcase, where our main interests lie, it can be given
the following strange interpretation. If we allow the real Minkowski space
to be complexified so that the real Minkowski coordinates x$^{a}$ take
complex values, i.e., x$^{a}\rightarrow z^{a}=x^{a}+i$y$^{a}$ with complex
metric $g=\eta _{ab}d$z$^{a}d$z$^{b},$ the real vacuum Maxwell equations can
be extended into the complex and rewritten as 
\begin{eqnarray*}
curlW &=&i\dot{W}, \\
divW &=&0 \\
\text{with }W &=&E+iB.
\end{eqnarray*}
This subcase of Maxwell fields can then be extended into the complex so as
to have as source, a complex analytic world-line, i.e., to now become \emph{%
\ \ complex} :Lienard-Wiechart fields. When viewed as real fields on the
real Minkowski space, ($z^{a}=x^{a}$), they possess a real principle null
vector that is shear-free but twisting and diverging. The twist is a measure
of how far the complex world-line is from the real 'slice'. Most Maxwell
fields in this subcase are asymptotically flat with a time-varying set of
electric and magnetic moments, all depending on the complex displacements
and the complex velocities.
\end{abstract}

\section{Introduction}

The study of null geodesics and null geodesic congruences, in physical
space-time, has played a large role in many branches of mathematical
physics, from the practical uses of the high frequency limit of
electromagnetic waves (leading to geometric optics and solutions of the
eikonal equation) to the beautiful study of the special solutions of the
Einstein Equations (the so-call algebraically special vacuum metrics) which
are determined by a special class of null geodesic congruences\cite
{R.T.,KerrEtc,Tal,Lind,Ivor,AT,PS}. This work is devoted to a study of the
physical meaning of this special class in the case of Minkowski space; a
future work will be devoted to their meaning in curved spaces. In the flat
space case, the physical meaning or interpretation will come via certain
associated Maxwell fields .

The special class being considered are those null geodesic congruences whose
null tangent vector field, [say $l^{a}(x^{b})$], is diverging and
shear-free. They could be either twisting\cite{Tal,Lind} or non-twisting\cite
{R.T.}. For many years these congruences were close to the center of
research activities in GR. Via the Goldberg-Sachs theorem\cite{G.S.} they
defined the algebraically special vacuum metrics; they gave rise to a large
class of exact solutions of the Einstein equations (e.g., the
Robinson-Trautman class, the Kerr-Schild class and many isolated metrics)
plus years of fruitless man-hour searches for general twisting vacuum
metrics); they were an early catalyst for Penrose's twistor theory; they led
to, probably, the earliest physical use of C-R manifold theory\cite{PS,LNT}.
But almost all the studies were to elucidate the mathematical properties of
these congruences. We know of little or virtually no discussion \{aside from
the case of the Kerr\cite{Kerr} and charged Kerr metric\cite{etal,NL,N}\} of
the physical origin or meaning of these congruences. In the special case of
these congruences when they are \textit{\ not} surface forming (the twisting
congruences), the physical meaning of the twist itself has posed something
of a mystery. We must stress that we are not saying that the geometric
structure of the congruences posed a mystery - their geometry is well
understood. We are asking instead what physical consequences can be
associated with the twisting congruences. Here we will study these
congruences and attempt to give a physical interpretation of them (and in
particular to the twist) in the context of Maxwell fields in flat
space-time. Though their analysis is not simple, we will see that they
generate rather remarkable solutions to the vacuum Maxwell equations. In the
conclusion, we will discuss their possible meaning in GR.

The Maxwell fields will be connected or associated with these null geodesic
congruences in the following way: we will require that one (of the two) real
null eigenvectors, $l^{a},$ of the Maxwell field, [by definition, a
principal null vector (pnv) of the Maxwell field], i.e., 
\begin{equation}
F_{ab}l^{b}=\lambda l_{a},  \label{pnv}
\end{equation}
be the tangent vector to a null, diverging, shear-free geodesic congruence.

It has long been known that such congruences break up into two classes;

$\bullet $ Those whose null tangent vectors are proportional to a gradient
field, 
\begin{equation}
l_{a}=\alpha \partial _{a}\varphi ,  \label{twistfree}
\end{equation}
i.e., are \textit{twist-free},

\[
l_{[a}\partial _{b}l_{c]}=0, 
\]

$\bullet $ and those that are twisting, 
\begin{equation}
l_{[a}\partial _{b}l_{c]}\neq 0.  \label{twisting}
\end{equation}

The twisting class itself breaks into a variety of subcases, namely

$\bullet $$\bullet $ there are those whose caustics remain in a spatially
finite region. We will refer to these as ``regular'' twisting congruences.
They are the ones of most interest for us.

$\bullet $$\bullet $ and those whose caustics extend spatially to infinity,
the ``singular'' twisting congruences.

$\bullet $$\bullet $ and those without any caustics. They appear, however,
to have line(s) of infinite twist and seem to lead to Maxwell fields with
line singularities extending to infinity. Until they are better understood,
we will give them only passing mention.

It is easy to find a large subclass of the regular twisting congruences,
though \thinspace it is not known if we have found all of them. However, for
simplicity, at the present we will refer to this subclass as the regular
twisting congruences.

Our main interest lies in the both the twist free and the regular twisting
cases. Little further mention will be made of the singular congruences.

The Maxwell equations can be explicitly integrated in these cases. The
needed data for a solution is surprisingly simple with clear physical
meaning to the data. We point out (the already known fact) that in the
twist-free case, the Maxwell fields are given by the value of the electric
charge, $e$, plus the free-choice of a time-like world-line in Minkowski
space. These solutions are simply the well-known Wienard-Liechert (L.W.)
fields, the radiation fields of an arbitrarily moving charged particle in
Minkowski space. For the regular twisting case, the Maxwell fields have a
much richer structure that has a strange but similar interpretation. Though
the fields are \textit{real} Maxwell fields on \textit{real} Minkowski space
they can be interpreted as the restriction to the real subspace of \textit{\
complex} Maxwell fields living on \textit{complex} Minkowski space, where
the data is again the value of the electric charge, $e$, but in this case,
the time-like world-lines of the (real) Lienard-Wiechert fields are replaced
by complex analytic time-like world-lines in complex Minkowski space. The
twist of the congruence is a measure of how far into the complex the
world-line extends. The physical consequences of this twist in terms of the
Maxwell field is that the Maxwell field now possesses an \textit{\ intrinsic}
magnetic-dipole moment. The magnetic dipole is determined (in a precise
sense) by how far the complex world-line is from the real part of complex
Minkowski space. We will refer to these `twisting' Maxwell Fields as \textit{%
\ \ \ complex} Lienard-Wiechert fields. Though they are the direct analogues
of the real L.W. fields, in the real case, at any one moment, the electric
dipole\thinspace moment can be transformed away by a real origin shift and
the world-line's four-velocity can be equated with the time direction by a
boost, however the magnetic dipoles created this way in the complex case,
are intrinsic. Neither complex translations nor complex boosts are
physically allowed to eliminate them.

\begin{remark}
We could have included in the definition of \textit{complex}
Lienard-Wiechert fields, world-lines with a zero-norm tangent vector
(complex null world lines). They however all appear to have wire
singularities going to spacial infinity.
\end{remark}

In Sec. II, we review the general description (in Minkowski space) of null
geodesic congruences and their specialization, via the optical parameters,
to those that are diverging and shear free. In addition, by integration of
their defining equation, their realization is explicitly displayed. One sees
easily, from the explicit expressions, how they split into the two
subclasses of twist-free and twisting.

In Sec. III, after reviewing the meaning of `principal null vectors (pnv)',
we discuss the extension of the real Maxwell fields and equations into
complex Minkowski space. In particular, we point out and stress the
Lienard-Wiechert fields and the complex generalization to \textit{complex}
Lienard-Wiechert fields. Using the properties of the shear-free congruences,
the Maxwell equations can be integrated explicitly. Unfortunately the
integration is done using the special coordinate system that is associated
with the congruence. Because of the unusual (unfamiliar) coordinates, the
physical meaning of the parameters in the solution is not immediately
obvious. For this reason, we initially simply state our main results
concerning the physical meaning of these solutions and postpone their proofs
until the following sections.

In Sec.IV, using the spin-coefficient version of Maxwells equations written
in the shear-free null coordinate and tetrad system, we integrate them with
the assumption that a principal null vector is the tangent vector to a
shear-free (regular) null geodesic congruence. In the process we point out
that they can indeed be interpreted as complex L.W. fields. Still their
physical interpretation is obscure.

In Sec. V we describe the transformation to a Bondi-like coordinate system
(a coordinate system based on future null cones with apex at the spatial
origin) from which one obtains the physical meaning of the Maxwell data,
(which in turn arose from the properties of the congruence) in terms of the
electric and magnetic moments. One sees that the magnetic moment is
associated with the complex position of the complex world line.

In Sec. VI, we discuss and conjecture how these results and their physical
meaning could extend to the vacuum Einstein and Einstein-Maxwell equations.

[Note that for typographical reasons we have used $\frak{d}$ in place of the
operator edth.]

\section{\protect\large Null Geodesic Congruences in Minkowski Space}

\subsection{null geodesic congruences}

An arbitrary null geodesic congruence in Minkowski space can be described by
the following analytic expression: 
\begin{equation}
x^{a}=Ut^{a}-\overline{L}m^{a}-L\overline{m}^{a}+(r-r_{0})l^{a}
\label{nullgeodesics}
\end{equation}
where the individual geodesics are labeled (or parametrized) by ($U$,$\zeta $
,$\overline{\zeta }$) and $r$ is an affine parameter along each geodesic. $%
L=L(U,\zeta ,\overline{\zeta })$ is an arbitrary complex function of the
geodesic parameters, $t^{a}$ is a time-like vector (normalized for
convenience so that $t^{a}t_{a}=2$), $l^{a}=l^{a}$($\zeta $,$\overline{\zeta 
}$) is the tangent null vector to the geodesics and parametrized by all
points on the $(\zeta $,$\overline{\zeta })$ sphere, $S^{2},$ $m^{a}+%
\overline{m}^{a}$ and $i$($m^{a}-\overline{m}^{a})$ are the two space-like
vectors (also parametrized by $(\zeta $,$\overline{\zeta })$) orthogonal to $%
t^{a}$ and $l^{a}.$ The function $r_{0}=r_{0}(U,\zeta ,\overline{\zeta })$
describes the arbitrary origin of the affine parameter $r.$ It is usually
chosen to simplify the analytic form of the divergence of the congruence.

The expression Eq.(\ref{nullgeodesics}) for the null geodesic congruence can
be understood by the following construction; (1) First the function $%
L=L(U,\zeta ,\overline{\zeta })$ is chosen, then (2) the values of the
parameters ($U$,$\zeta $,$\overline{\zeta }$) are selected. (3) Starting
from the coordinate origin move along the time axis, $t^{a},$ a `distance' $%
U,$ to the point, ($x^{a}=Ut^{a}$), then (4), [with the chosen $(\zeta $,$%
\overline{\zeta }),$ (and thus, from $l^{a}$($\zeta $,$\overline{\zeta }$),
a null direction on the $S^{2}$ of null directions)] move perpendicular to ($%
t^{a}$, $l^{a})$ to the point $x^{a}=Ut^{a}-\overline{L}m^{a}-L\overline{m}
^{a}$ and finally, (5), at that point construct the null geodesic by moving
in the direction of the tangent vector $l^{a}(\zeta ,\overline{\zeta }),$ an
arbitrary `distance' $r.$

\textit{Any} null geodesic congruence can be selected in this manner by the
appropriate choice of the arbitrary $L(U,\zeta ,\overline{\zeta }).$

Eq.(\ref{nullgeodesics}) has a dual interpretation. In addition to
describing an arbitrary null geodesic congruence, it can be viewed as a
coordinate transformation from standard Minkowski coordinates, $x^{a},$ to
the null geodesic coordinates, $(U,r,\zeta ,\overline{\zeta }).$ This
alternate point of view is important to us, since the integration of the
Maxwell equations with our conditions, is quite simple in these coordinates
but almost impossible in the standard coordinates.

For later use, we define the full null-tetrad (parametrized by $(\zeta $,$%
\overline{\zeta })$) 
\begin{eqnarray}
\lambda _{i}^{a} &=&(l^{a},n^{a},m^{a},\overline{m}^{a}),  \label{nullTetrad}
\\
n^{a} &=&t^{a}-l^{a}  \nonumber
\end{eqnarray}

\noindent where

\[
l^{a}n_{a}=1=-m^{a}\overline{m}_{a},\text{ \quad all other products
vanishing.} 
\]

\begin{remark}
Often it is useful to have an explicit representation of the tetrad; 
\begin{eqnarray}
l^{a} &=&\frac{\sqrt{2}}{2}(1,\frac{\zeta +\overline{\zeta }}{1+\zeta 
\overline{\zeta }},-i\frac{\zeta -\overline{\zeta }}{1+\zeta \overline{\zeta 
}},\frac{-1+\zeta \overline{\zeta }}{1+\zeta \overline{\zeta }});
\label{tetrad} \\
m^{a} &=&\text{$\frak{d}$}l^{a}=\frac{\sqrt{2}}{2}(0,\frac{1-\overline{\zeta 
}^{2}}{1+\zeta \overline{\zeta }},\frac{-i(1+\overline{\zeta }^{2})}{1+\zeta 
\overline{\zeta }},\frac{2\overline{\zeta }}{1+\zeta \overline{\zeta }}) 
\nonumber \\
\overline{m}^{a} &=&\overline{\text{$\frak{d}$}}l^{a}=\frac{\sqrt{2}}{2}(0,%
\frac{1-\zeta ^{2}}{1+\zeta \overline{\zeta }},\frac{i(1+\zeta ^{2})}{%
1+\zeta \overline{\zeta }},\frac{2\zeta }{1+\zeta \overline{\zeta }}) 
\nonumber \\
n^{a} &=&t^{a}-l^{a}=\frac{\sqrt{2}}{2}(1,-\frac{\zeta +\overline{\zeta }}{%
1+\zeta \overline{\zeta }},i\frac{\zeta -\overline{\zeta }}{1+\zeta 
\overline{\zeta }},\frac{1-\zeta \overline{\zeta }}{1+\zeta \overline{\zeta }%
}).  \nonumber
\end{eqnarray}
The vectors in this representation are parametrized by ($\zeta ,\overline{%
\zeta }$) but the components are given in the Minkowski coordinate system.
They can be represented in the null geodesic coordinates, $(U,r,\zeta ,%
\overline{\zeta })$ via the coordinate transformation, Eq.(\ref
{nullgeodesics}).
\end{remark}

\subsection{The optical parameters;\textbf{\ divergence, shear and twist}}

The optical parameters will be described in the language and notation of
spin-coefficients.

The quantity $\rho $ is referred to as the complex divergence and defined by

\begin{eqnarray}
\rho &=&m^{a}\overline{m}^{b}\nabla _{a}l_{b}=\frac{1}{2}(-\nabla
_{a}l^{a}+icurl\,l^{a})  \label{div} \\
curl\,\,l^{a} &=&(\nabla _{[a}l_{b]}\nabla ^{a}l^{b})^{\frac{1}{2}} 
\nonumber
\end{eqnarray}
with the shear $\sigma $ defined by 
\begin{equation}
\sigma =m^{a}m^{b}\nabla _{a}l_{b}.  \label{shear}
\end{equation}

By integrating the optical equations (assuming that $\rho \neq 0$) 
\begin{eqnarray*}
DP &\equiv &\partial _{r}P=P^{2} \\
P &=&\left\| 
\begin{array}{cc}
\rho , & \sigma \\ 
\overline{\sigma }, & \overline{\rho }
\end{array}
\right\|
\end{eqnarray*}
one finds 
\begin{eqnarray}
\rho &=&\frac{i\Sigma -r}{r^{2}+\Sigma ^{2}-\overline{\sigma }_{0}\sigma _{0}%
}  \label{rhosigma} \\
\sigma &=&\frac{\sigma _{0}}{r^{2}+\Sigma ^{2}-\overline{\sigma }_{0}\sigma
_{0}}  \nonumber
\end{eqnarray}
with ($\Sigma ,\sigma _{0}$) unknown (integration) functions of $(U,\zeta ,%
\overline{\zeta }).$ The origin of $r$ has already been chosen to make $%
\Sigma $ real.

By a direct (but fairly lengthy) calculation from Eq.(\ref{nullgeodesics})
and comparison with Eq.(\ref{rhosigma}) we have that 
\begin{eqnarray}
\sigma _{0} &=&\text{$\frak{d}$}L+LL,_{{\small U}}  \label{sgma0} \\
2i\Sigma &=&\text{$\frak{d}$}\overline{L}+L\overline{L},_{{\small U}}-\text{ 
}\overline{\text{$\frak{d}$}}L-\overline{L}L,_{{\small U}}  \label{SIGMA} \\
r_{0} &=&-\frac{1}{2}(\text{$\frak{d}$}\overline{L}+L\overline{L},_{{\small U%
}}+\overline{\text{$\frak{d}$}}L+\overline{L}L,_{{\small U}}).\text{ }
\label{r0}
\end{eqnarray}
We thus see that the optical parameters are determined by our choice of $%
L(U,\zeta ,\overline{\zeta }).$

If we now require that the congruence be shear-free, from Eq.(\ref{sgma0}),
we have the differential condition on $L$

\begin{equation}
\sigma _{0}=\text{$\frak{d}$}L+LL,_{{\small U}}=0  \label{shearfree}
\end{equation}
so that the divergence becomes

\begin{equation}
\rho =-\frac{1}{(r+i\Sigma )}.  \label{rho}
\end{equation}

The $\Sigma $ is, thus, a measure of the twist (curl). For simplicity we
will refer to $\Sigma $ as the twist.

The issue now is how to find the solutions to Eq.(\ref{shearfree}) for $%
L(U,\zeta ,\overline{\zeta })$.

\begin{remark}
It should be emphasized that much of the material of this section is not
new. It has been developed by many workers and has appeared in many
versions. For example, finding solutions (see below) to Eq.(\ref{shearfree})
was an early result in Penrose's Twistor theory\cite{RP} and is often
referred to as the Kerr theorem.
\end{remark}

Our approach to Eq.(\ref{shearfree}) is based on the observation that by the
introduction of the (complex) potential for $L$, 
\begin{equation}
\phi =\phi (U,\zeta ,\overline{\zeta })  \label{phi}
\end{equation}
via 
\begin{equation}
\text{$\frak{d}$}\phi +L\,\phi ,_{{\small U}}=0,  \label{Lofphi}
\end{equation}
the analysis is greatly simplified. One, however, must be aware that later
we encounter interesting difficulties with the meaning of the \textit{complex%
} $\phi .$

Direct substitution of 
\begin{equation}
L\,=-\frac{\text{$\frak{d}$}\phi }{\phi ,_{{\small U}}}  \label{L1}
\end{equation}
into Eq.(\ref{shearfree}) does not simplify at all but if we reverse the
point of view and \textit{treat} $\phi $ as the \textit{independent variable}
so that Eq.(\ref{phi}) is rewritten implicitly as

\begin{equation}
U=X(\phi ,\zeta ,\overline{\zeta })  \label{U}
\end{equation}
a huge simplification occurs.

From Eq.(\ref{U}) we have 
\begin{eqnarray}
1 &=&X_{\phi }(\phi ,\zeta ,\overline{\zeta })\,\phi ,_{{\small U}}
\label{XDerivates} \\
0 &=&X_{\phi }(\phi ,\zeta ,\overline{\zeta })\text{$\frak{d}$}\phi +\text{$%
\frak{d}$}_{|\phi }X  \nonumber
\end{eqnarray}
\{$\frak{d}_{|\phi }$ means the edth derivative holding $\phi $ constant
rather than holding $U$ constant\} so that 
\begin{equation}
L\,=-\frac{\text{$\frak{d}$}\phi }{\phi ,_{{\small U}}}=\text{$\frak{d}$}%
_{|\phi }X.  \label{LofX}
\end{equation}

\begin{remark}
For any given $\phi =\phi (U,\zeta ,\overline{\zeta }),$ we see that $\phi
^{*}=\Phi (\phi ,\overline{\zeta }),$ with arbitrary $\Phi (\phi ,\overline{%
\zeta }),$ yields the same $L,$ since 
\begin{eqnarray}
\text{$\frak{d}$}\phi ^{*} &=&\partial _{\phi }\Phi \cdot \text{$\frak{d}$}%
\phi ^{*}  \label{reparametrize} \\
\phi ^{*},_{{\small U}} &=&\partial _{\phi }\Phi \,\cdot \,\phi ,_{{\small U}%
}  \nonumber
\end{eqnarray}
and 
\[
\frac{\text{$\frak{d}$}\phi ^{*}}{\phi ^{*},_{{\small U}}}=\frac{\text{$%
\frak{d}$}\phi }{\phi ,_{{\small U}}}. 
\]

This freedom in the choice of $\phi ,$ which is quite useful later, is
referred to as the `regauging' of $\phi $.
\end{remark}

Using Eqs.(\ref{LofX}) and (\ref{XDerivates}), Eq.(\ref{shearfree}) becomes
the easily integrable equation

\begin{equation}
\text{$\frak{d}$}_{|\phi }^{2}X=0  \label{edth2X}
\end{equation}
whose general solution has the form 
\begin{equation}
u=X=\frac{a(\phi ,\overline{\zeta })+b(\phi ,\overline{\zeta })\zeta }{%
1+\zeta \overline{\zeta }}=\frac{a(\phi ,\overline{\zeta })+\phi \zeta }{%
1+\zeta \overline{\zeta }}  \label{sol}
\end{equation}
where the regauging freedom allowed $b(\phi ,\overline{\zeta })=>\varphi .$
Using (\ref{LofX}) we have that

\begin{equation}
L=\frac{\phi -a(\phi ,\overline{\zeta })\overline{\zeta }}{1+\zeta \overline{
\zeta }}  \label{L*}
\end{equation}
from which it follows that

\begin{eqnarray*}
L+u\overline{\zeta } &=&\phi \\
u-L\zeta &=&a(\phi ,\overline{\zeta }).
\end{eqnarray*}
By eliminating the $\varphi ,$ we see that the general solution for $L$($%
u,\zeta \overline{,\zeta },$) is given by an arbitrary function 
\begin{equation}
F(u-L\zeta ,L+u\overline{\zeta },\overline{\zeta })=0.  \label{twistor}
\end{equation}
This, the Kerr theorem, is an important result in twistor theory; shear
free-null geodesic congruences are described by arbitrary analytic functions
of three variables, ($u-L\zeta ,$ $L+u\overline{\zeta },$ $\overline{\zeta }
).$

Returning to Eq.(\ref{sol}), we point out that for most choices of $a(\phi ,%
\overline{\zeta })$ the solution $X(\phi ,\zeta ,\overline{\zeta })$ will
have angular singularities. The only choices that avoid them are 
\[
a(\phi ,\overline{\zeta })=\alpha (\phi )+\beta (\phi )\overline{\zeta }. 
\]
Dropping the regauging condition, $b(\phi ,\overline{\zeta })=>\phi ,$ we
have that the general non-singular solution to (\ref{edth2X}) is

\begin{equation}
U=X(\phi ,\zeta ,\overline{\zeta })=\frac{\alpha (\phi )+\beta (\phi )%
\overline{\zeta }+\overline{\beta }(\phi )\zeta +\gamma (\phi )\zeta 
\overline{\zeta }}{1+\zeta \overline{\zeta }}.  \label{non-sing}
\end{equation}

In terms of a spherical harmonic expansion of $X(\phi ,\zeta ,\overline{
\zeta }),$ this expression involves just the $(l=0,1)$ terms with the four
coefficients given as arbitrary complex analytic functions of $\phi .$ Note
that the four components of $l_{a}(\zeta ,\overline{\zeta })$ are linear
combinations of the four $(l=0,1)$ harmonics which leads to the concise (and
useful) expression, 
\begin{equation}
U=X(\phi ,\zeta ,\overline{\zeta })=\xi ^{a}(\phi )l_{a}(\zeta ,\overline{
\zeta })  \label{sol2}
\end{equation}

\begin{definition}
We will consider only these non-singular solutions. The null geodesic
congruences obtained from the non-singular solutions, even though they, in
general, have caustics, will be referred to as non-singular congruences. If
the ``velocity vector'', $\partial _{\phi }\xi ^{a}(\phi ),$ has a non-zero
norm, 
\[
\partial _{\phi }\xi ^{a}(\phi )\partial _{\phi }\xi ^{b}(\phi )\eta
_{ab}\neq 0, 
\]
we will refer to the congruence as a regular congruence.
\end{definition}

\begin{remark}
The caustics (where $\rho =\infty $) are explicitly constructed from Eqs.(%
\ref{rho}) and (\ref{nullgeodesics}) by setting $r=0$ and $\Sigma (U,\zeta ,%
\overline{\zeta })=0.$ We have 
\begin{equation}
x^{a}=Ut^{a}-\overline{L}(U,\zeta ,\overline{\zeta })m^{a}-L(U,\zeta ,%
\overline{\zeta })\overline{m}^{a}-r_{0}(U,\zeta ,\overline{\zeta })l^{a}
\label{caustic}
\end{equation}

and 
\[
\Sigma (U,\zeta ,\overline{\zeta })=0. 
\]
At a fixed time, $x^{0},$we have from $x^{0}=x^{0}(U,\zeta ,\overline{\zeta }%
)$ an expression of the form, $U=C(x^{0},\zeta ,\overline{\zeta }).$ This,
with $\Sigma (U,\zeta ,\overline{\zeta })=0,$ yields, in general,
expressions 
\begin{eqnarray*}
\zeta &=&Z(U,x^{0}) \\
\overline{\zeta } &=&\overline{Z}(U,x^{0})
\end{eqnarray*}
so that the caustic, at fixed time $x^{0}$, is a curve, a `string', given
parametrically by 
\[
x^{i}=\chi ^{i}(U,x^{0}). 
\]

As $x^{0}$ changes, this describes parametrically a two-surface (a `string'
evolving in time) in the Minkowski space. For many choices of the $\xi
^{a}(\phi ),$ the `strings' are closed.
\end{remark}

\begin{remark}
The complex conjugate $\phi ,$ i.e., $\overline{\phi }=$ $\overline{\phi }%
(U,\zeta ,\overline{\zeta })$ is obtained implicitly from 
\begin{equation}
U=\overline{\xi }^{a}(\overline{\phi })l_{a}(\zeta ,\overline{\zeta }).
\end{equation}
\end{remark}

Later we will interpret the $\xi ^{a}(\phi )$ as the parametric description
a \textit{complex} world-line in \textit{complex} Minkowski space by

\begin{equation}
z^{a}=\xi ^{a}(\phi ),  \label{compexW.L.}
\end{equation}
with $z^{a}$ the complex Minkowski coordinates. In the case where the norm
of the ``velocity vector'', $v^{a}\equiv \partial _{\phi }\xi ^{a},$ is
different from zero the regauging of $\phi $ allows us to choose $\phi $ so
that $\eta _{ab}v^{a}v^{b}=1.$ In the future we will make this choice. The
case of zero norm is actually quite interesting and includes the very
important Robinson null geodesic congruence. We are most interested in the
case of the non-zero norms.

A special solution of Eq.(\ref{edth2X}) (not in general included in Eq.(\ref
{non-sing}) is to have the four functions $\xi ^{a}(\phi )$ be real
functions of a real $\phi _{re},$ with no assumptions of analyticity. They
can then be interpreted as the parametric form of a real world-line in real
Minkowski space,

\begin{eqnarray}
x^{a} &=&\xi ^{a}(\tau ).  \label{realW.L.} \\
\tau &=&\phi _{re}
\end{eqnarray}
When the $\xi ^{a}(\tau )$ are real analytic functions this is the special
case of Eq.(\ref{complexW.L.}). This case is determined by having real
coefficients for the Taylor series of the $\xi ^{a}(\phi )$ and choosing
real $\phi .$

Using Eq.(\ref{non-sing}), $L$ is found, from Eq.(\ref{LofX}), to be 
\begin{equation}
L=\xi ^{a}(\phi )m_{a}(\zeta ,\overline{\zeta }).  \label{L}
\end{equation}

Calculating the twist from Eqs.(\ref{L}) and (\ref{SIGMA}) we obtain 
\begin{equation}
2i\Sigma =\{\xi ^{a}(\varphi )-{}\overline{\xi }^{a}(\overline{\varphi }
)\}\{t_{a}-m_{a}\overline{L},_{{\small U}}-\overline{m}_{a}L,_{{\small U}}\}
\label{SIGMA2}
\end{equation}
from which we immediately see that in the case of a real world-line, we have
that $\Sigma =0,$ i.e. the twist vanishes, while for an intrinsically
complex world-line the twist is non-zero$.$ We thus have our two classes of
shear-free diverging null geodesic congruences.

There is a \textit{very important observation} to be made concerning the
equations for the non-singular shear-free null geodesic congruences,

\begin{eqnarray}
x^{a} &=&Ut^{a}-\overline{L}m^{a}-L\overline{m}^{a}+(r-r_{0})l^{a},
\label{nullgeodesics2} \\
L(U,\zeta ,\overline{\zeta }) &=&\xi ^{a}(\phi )m_{a}(\zeta ,\overline{\zeta 
}),  \label{L2} \\
U &=&\xi ^{a}(\phi )l_{a}(\zeta ,\overline{\zeta }).  \label{U2}
\end{eqnarray}
By allowing both sets of the coordinates $x^{a}$ and ($U,r,\zeta ,\overline{%
\zeta }$) to take on complex values (which is allowed since the expressions
are all analytic), the Eq.(\ref{nullgeodesics2}) can be written as 
\begin{equation}
z^{a}=\xi ^{b}(\phi )l_{b}(\zeta ,\overline{\zeta })t^{a}-\overline{\xi }%
^{b}(\overline{\phi })\overline{m}_{b}(\zeta ,\overline{\zeta })m^{a}-\xi
^{b}(\phi )m_{b}(\zeta ,\overline{\zeta })\overline{m}^{a}+(r-r_{0})l^{a}.
\label{ng4}
\end{equation}
By expanding $\xi ^{b}(\phi )$ in the tetrad basis, ($l^{a},m^{a},\overline{m%
}^{a},n^{a}$), this becomes 
\begin{equation}
\text{ }z^{a}=\xi ^{a}(\phi )-Q^{0}m^{a}+(r+i\Sigma
+\{Q^{0}L_{U}\})l^{a}(\zeta ,\overline{\zeta })  \label{ng5}
\end{equation}
with 
\[
Q^{0}=\frac{\overline{\text{$\frak{d}$}}\phi }{\phi ,_{{\small U}}}+%
\overline{L}. 
\]
Finally using 
\begin{eqnarray*}
V &\equiv &\xi ^{a},_{\phi }l_{a}(\zeta ,\overline{\zeta })=\frac{1}{\phi ,_{%
{\small U}}} \\
Q &\equiv &\rho Q^{0}
\end{eqnarray*}
and introducing new variables 
\begin{eqnarray}
r^{*} &=&(r+i\Sigma )V  \label{r*} \\
\phi &=&\phi (U,\zeta ,\overline{\zeta })\Leftrightarrow U=\xi ^{a}(\phi
)l_{a}(\zeta ,\overline{\zeta })  \label{phi*} \\
\{\zeta ^{*},\overline{\zeta }^{*}\} &=&\{\frac{\{\zeta +\frac{Q}{(1-QL_{U})}%
\}}{[1-\frac{Q\overline{\zeta }}{(1-QL_{U})}]},\text{ }\overline{\zeta }\}
\label{zetab*}
\end{eqnarray}
and new null vector 
\begin{eqnarray}
l^{*a} &=&V^{-1}(1-QL,_{{\small U}})l^{a}+V^{-1}Qm^{a},  \label{newl} \\
l^{*a}(\zeta ^{*},\overline{\zeta }^{*},\phi ) &=&V^{*-1}l^{a}(\zeta ^{*},%
\overline{\zeta }^{*})\equiv V^{*-1}\frac{\sqrt{2}}{2}(1,\frac{\zeta ^{*}+%
\overline{\zeta }^{*}}{1+\zeta ^{*}\overline{\zeta }^{*}},-i\frac{\zeta ^{*}-%
\overline{\zeta }^{*}}{1+\zeta ^{*}\overline{\zeta }^{*}},\frac{-1+\zeta ^{*}%
\overline{\zeta }^{*}}{1+\zeta ^{*}\overline{\zeta }^{*}}) \\
V^{*} &=&\frac{V}{(1-QL_{U})}
\end{eqnarray}
Eq.(\ref{ng5}) becomes 
\begin{equation}
z^{a}=\xi ^{a}(\phi )+r^{*}l^{*a}(\zeta ^{*},\overline{\zeta }^{*},\phi ).
\label{complexl.c.}
\end{equation}

There are two extremely important points to be made from this transformation;

1. Eq.(\ref{complexl.c.}) describes a \textit{complex} shear-free null
geodesic congruence where the null geodesics are the generators of the
complex light-cones having apex on the complex world-line $\xi ^{a}(\phi ).$
This congruence is \textit{twist-free} simply by construction. Almost all
the null geodesics are complex; some of them however pass through
(intersect) the real Minkowski space.

2. From Eq.(\ref{newl}) we see that by projecting the complex $l^{*a}$ into
the real Minkowski space, i.e., deleting the term proportional to the
(complex) $m^{a},$ we obtain our twisting null geodesic tangent field $%
l^{a}(x^{a}).$ The \textit{real} shear-free twisting null tangent field of
our earlier discussion can be viewed as the \textit{real projection} of a 
\textit{twist-free complex tangent field} (obtained from the complex
light-cones with apex on the complex world-line) into the real Minkowski
space.

Though we will not use the details of this transformation, these
observations will play an important role in the interpretation of certain
Maxwell fields.

\section{\protect\large \ Maxwell Fields}

The principal null vectors (pnv), $l^{b}$\ of a Maxwell field are defined by
the eigenvector equation 
\begin{equation}
F_{ab}l^{b}=\lambda l_{a}.  \label{eigenvector}
\end{equation}
In general, at each space-time point, there are two independent real
eigenvectors (non-degenerate). (There is the degenerate case, which we will
not consider, when the two real null eigenvectors coincide.)

Our interest will center on the non-degenerate case.

Our goal is to solve the Maxwell equations with the condition that one of
the eigenvectors, $l^{b}(x^{a}),$ defines a null vector field that is the
tangent field of a \textit{diverging, shear-free null geodesic congruence}.
In this manner we can give physical meaning, via the Maxwell field, to the
variables associated with this type of congruence.

\subsection{Complex Maxwell Fields}

To begin with we will review\cite{cMF} the idea of the extension of the
Maxwell field and equations from real Minkowski space to complex Minkowski
space.

\begin{remark}
The idea and use of complex Maxwell fields has a long and illustrious
history; apparently in non-published notes Riemann first pointed out the
naturalness of the complexification. His observations and applications were
later followed by those of L. Silberstein\cite{Iwo}, C. Lanczos\cite{Gs} and
H. Bateman\cite{B}.
\end{remark}

First note that the real vacuum Maxwell equations 
\[
\nabla _{a}F^{ab}=0=\nabla _{a}F^{*ab} 
\]
can be written as 
\begin{eqnarray}
&&\nabla _{a}W^{ab}=0  \label{complexMax} \\
W^{ab} &=&F^{ab}+iF^{*ab}  \nonumber
\end{eqnarray}
\quad or 
\begin{eqnarray}
curl\,\vec{W} &=&i\partial _{t}\vec{W},\qquad div\text{ }\vec{W}=0,
\label{complexMax2} \\
&&\text{with }\vec{W}=\vec{E}+i\vec{B}.  \nonumber
\end{eqnarray}

If one deals with (piece-wise) real analytic fields, they can be
analytically continued into the complex Minkowski space, i.e., 
\begin{eqnarray}
&&W^{ab}(x^{a})\Rightarrow W^{ab}(z^{a})  \label{analyticExt} \\
z^{a} &=&x^{a}+iy^{a}.  \nonumber
\end{eqnarray}
Likewise the Maxwell equations, Eq.(\ref{complexMax}) or (\ref{complexMax2})
are easily extended into the complex by the simple substitution, in the
derivatives, of $x^{a}\Rightarrow z^{a}.$ We can thus work with complex
Maxwell fields in complex Minkowski space.

This complexification process can be easily reversed: From a complex
analytic solution, $W^{ab}(z^{a}),$ one obtains a real solution by letting $%
z^{a}=x^{a}$ and then taking the real and imaginary parts of $W^{ab}(z^{a}),$
i.e., ($F^{ab},F^{*ab}),$ (or of $\vec{W}$ to obtain $\vec{E}$ and $\vec{B}$
). We are thus dealing with complex Maxwell fields on complex Minkowski
space that can be understood from the point of view of real fields on real
Minkowski space or alternatively the real fields can be understood from the
complex point of view. It is this latter idea that we will explore.

(Aside: There is reverse process that is actually richer than we need here.
From any complex $W^{ab}(z^{a})$ we can restrict $z^{a}$ to $%
z^{a}=x^{a}+i\alpha ^{a}$ with constant $\alpha ^{a}$ and then take the real
and imaginary parts of $W^{ab}.$ This yields a four parameter family of real
Maxwell fields from a single complex one. Further generalizations do exist.)

An important class of real vacuum Maxwell fields are the Lienard-Wiechert
fields\cite{LL}, the radiation fields from a moving electric monopole on an
arbitrary (real) time-like world line. It is now easy to consider their
generalization to \textit{complex} Lienard-Wiechert fields, i.e., to
consider complex Maxwell fields whose source is an \textit{electric} charge
moving on a \textit{complex} analytic world-line in complex Minkowski space.
The idea will be to find these solutions and then interpret the real fields
in terms of the complex world line.

Since the details of the derivations and proofs are relatively complicated
and long, we will first state the results and give the proofs afterwards.

\subsection{Results}

The first result, which concerns the real Lienard-Wiechert solution, has
already been known\cite{LW1,LW2} for many years.

\textbf{Theorem I}:

A Maxwell field is a (real) Lienard-Weichert field if and only if a pnv of
the field is the tangent vector to a diverging, shear-free and \textit{non}-%
\textit{twisting} null geodesic congruence. The charge moves on the real
world line, $x^{a}=\xi ^{a}(\tau ),$ Eq.(\ref{realW.L.}), that is associated
with the congruence.

\begin{definition}
The world-line $x^{a}=\xi ^{a}(\tau )$ is referred to as the (real) center
of charge world-line.
\end{definition}

\textbf{Corollary}; In the (instantaneous) rest frame, i.e., with $%
v^{a}=\partial _{\tau }\xi ^{a}(\tau )=\sqrt{2}t^{a}=\sqrt{2}\delta
_{0}^{a}, $ the radiation field is pure dipole, $q\partial _{\tau }^{2}\xi
^{a}(\tau )\cdot \overline{m}_{a}(\zeta ,\overline{\zeta }),$ and the
electric dipole moment is $d^{a}=q[\xi ^{a}(\tau )-\frac{1}{2}\xi
^{b}t_{b}t^{a}].$ At any one instant $d^{a}$ can be put to zero by a spatial
origin shift.

\textbf{Theorem II: }

If the pnv of a Maxwell field is the tangent field of a (real) diverging,
shear-free and \textit{twisting, non-singular} null geodesic congruence then
the \textit{real} Maxwell field is derived from the complex Lienard-Wiechert
field by reversing the complexification, i.e., by looking at the real and
imaginary parts of $\overrightarrow{W}$ on the real Minkowski space. The
complex world-line associated with the Maxwell field coincides with the
complex world-line associated with the congruence, Eq.(\ref{complexW.L.}), $%
z^{a}=\xi ^{a}(\varphi )$.

\begin{definition}
The complex world-line $z^{a}=\xi ^{a}(\varphi )$ will be referred to as the 
\textit{complex center of charge world-line.}
\end{definition}

The complex world-line can be decomposed into real and imaginary parts;

\begin{equation}
\xi ^{a}(\varphi )=\xi _{R}^{a}(\varphi )+i\xi _{I}^{a}(\varphi ).
\label{complexW.L.2}
\end{equation}

They are defined from the real and imaginary parts of the coefficients of
the Taylor series for $\xi ^{a}(\varphi ).$ If $\xi _{I}^{a}(\varphi )$
vanishes we are back to the case of Theorem I.

\begin{example}
A simple but fairly general example can be constructed by taking a real
analytic world-line followed by a constant imaginary displacement; i.e.,
with $\xi _{0}^{a}$ a constant displacement, 
\begin{equation}
\xi ^{a}(\varphi )=\xi _{R}^{a}(\varphi )+i\xi _{0}^{a}.
\end{equation}
\end{example}

\begin{remark}
A caveat must be emphasized. The decomposition, Eq.(\ref{complexW.L.2}), is
regauging dependent and care must be used. The vanishing of $\xi
_{I}^{a}(\varphi )$ as really defining a gauge equivalence class.
\end{remark}

If $\xi _{I}^{a}(\varphi )$ is a constant space-like vector and if $\xi
_{R}^{a}(\varphi )$ were just a time-like linear function of $\varphi ,$ the
resulting Maxwell field is (what is referred to as) the Kerr-Maxwell field,
i.e., the Maxwell field found from the flat-space limit of the charged Kerr
metric\cite{etal,N}. In this latter case the Maxwell field is stationary and
asymptotically flat with a magnetic dipole moment (essentially) equal to the
charge $q,$ times the constant imaginary displacement $\xi _{I}^{a}.$

In special cases, as for example the Kerr solution, it has been possible to
analyze in detail the physical meaning of the complex world-line, however
for the general situation the full analysis has not yet been completed.
Nevertheless, one can say that in the cases with non-vanishing twist, the
twist (which via Eq.(\ref{SIGMA2})) is a measure of the imaginary
displacement $\xi _{I}^{a}(\varphi ),$ gives rise, in the real solution, to
a magnetic dipole and in general to magnetic dipole radiation. A reason this
analysis has proven to be difficult is that the solution for the Maxwell
field has been given in the spin-coefficient version of Maxwell's equations
using the tetrad associated with the null geodesic congruence, Eq.(\ref
{nullTetrad}) and the local congruence coordinates, $(U,r,\zeta ,\overline{
\zeta })$ and the transformation to more standard coordinates and tetrad is
quite complicated. In the next Section, to help understand the physical
meaning of these Maxwell fields, we will give the asymptotic version of this
transformation.

\section{Proof}

\subsection{The Maxwell Field}

The detailed proof of our theorems is fairly involved. It begins with the
spin-coefficient version of the Maxwell equations in the coordinate and
tetrad system associated with the shear-free twisting congruence. This has
already been found and partially integrated\cite{NL,Lind}. Rather than
repeat this calculation, we will simple quote the relevant parts. The final
integration is however not difficult and we are able to give a simple
expression for the complete Maxwell field. The difficulty lies in the
physical interpretation. In order to do this we will have to reexpress the
solution in a more physical coordinate system, namely a Bondi-system. (A
null coordinate system based on the future light-cones with apex on a
time-like world line at the spatial origin.)

The Maxwell equations, written in spin-coefficient notation, with 
\begin{eqnarray}
\phi _{0} &=&F_{ab}l^{a}m^{b}  \label{MaxField} \\
\phi _{1} &=&\frac{1}{2}F_{ab}(l^{a}n^{b}+m^{a}\overline{m}^{b})  \nonumber
\\
\phi _{2} &=&F_{ab}\overline{m}^{a}n^{b},  \nonumber
\end{eqnarray}
become 
\begin{eqnarray}
D\phi _{1}-\overline{\delta }\phi _{0} &=&(\pi -2\alpha )\phi _{0}+2\rho
\phi _{1}-\kappa \phi _{2},  \label{spmaxwell} \\
D\phi _{2}-\overline{\delta }\phi _{1} &=&-\lambda \phi _{0}+2\pi \phi
_{1}+(\rho -2\epsilon )\phi _{2},  \nonumber \\
\delta \phi _{1}-\bigtriangleup \phi _{0} &=&(\mu -2\gamma )\phi _{0}+2\tau
\phi _{1}-\sigma \phi _{2},  \nonumber \\
\delta \phi _{2}-\bigtriangleup \phi _{1} &=&-\upsilon \phi _{0}+2\mu \phi
_{1}+(\tau -2\beta )\phi _{2}.  \nonumber
\end{eqnarray}
with 
\[
D=l^{a}\partial _{a},\text{ }\delta =m^{a}\partial _{a},\text{ }\overline{%
\delta }=\overline{m}^{a}\partial _{a},\text{ }\bigtriangleup =n^{a}\partial
_{a}. 
\]
Using the spin coefficients, calculated in the coordinate system $(U,r,\zeta
,\overline{\zeta })$ and the associated null tetrad system\cite{NL,Lind},

\begin{eqnarray*}
\kappa &=&\sigma =\epsilon =\pi =\tau =\lambda =\gamma =0 \\
\rho &=&-(r+i\Sigma )^{-1} \\
\alpha &=&\frac{1}{2}\zeta \rho ;\text{ }\beta =-\frac{1}{2}\overline{\zeta }%
\overline{\rho } \\
\mu &=&-(1+\text{$\frak{d}$}\overline{L}_{U}+L\overline{L}_{UU})\overline{%
\rho } \\
&&v=\overline{L}_{UU},
\end{eqnarray*}
and the condition for $l^{a}$ to be an eigenvector (pnv) of $F_{ab},$ namely

\begin{equation}
\phi _{0}=0  \label{phi=0}
\end{equation}
we have\cite{NL,Lind}, from the first two of the Maxwell equations, Eqs.(\ref
{spmaxwell}), that

\begin{eqnarray}
\phi _{0} &=&0  \label{sol.I} \\
\phi _{1} &=&\rho ^{2}\phi _{1}^{0} \\
\phi _{2} &=&\rho \phi _{2}^{0}+\rho ^{2}\phi _{2}^{1}+\rho ^{3}\phi _{2}^{2}
\nonumber
\end{eqnarray}
with $(\phi _{1}^{0},\phi _{2}^{0},\phi _{2}^{1},\phi _{2}^{2})$ functions
of $(U$,$\zeta $,$\overline{\zeta }).$ The $(\phi _{2}^{1},\phi _{2}^{2}),$
though known expressions\cite{NL,Lind} in terms of ($\phi _{1}^{0},\phi
_{2}^{0}$)$,$ are not needed for any of the following.

The second two of Maxwell's equations, Eqs.(\ref{spmaxwell}) determine
equations for $(\phi _{1}^{0},\phi _{2}^{0}$), namely 
\begin{eqnarray}
\text{$\frak{d}$}\phi _{1}^{0}+L\cdot \partial _{{\small U}}\phi
_{1}^{0}+2\partial _{{\small U}}L\cdot \phi _{1}^{0} &=&0  \label{lastMaxEqs}
\\
\text{$\frak{d}$}\phi _{2}^{0}+L\cdot \partial _{{\small U}}\,\phi
_{2}^{0}+\partial _{{\small U}}L\,\cdot \phi _{2}^{0} &=&\partial _{{\small U%
}}\,\phi _{1}^{0}.  \nonumber
\end{eqnarray}
These equations, though looking apparently quite formidable, are actually
easily solved when $L$ is given by Eqs. (\ref{L2}) and (\ref{U2}); 
\begin{eqnarray}
L(U,\zeta ,\overline{\zeta }) &=&\xi ^{a}(\phi )m_{a}(\zeta ,\overline{\zeta 
}),  \label{L3} \\
U &=&X(\phi ,\zeta ,\overline{\zeta })=\xi ^{a}(\phi )l_{a}(\zeta ,\overline{%
\zeta }),  \label{U3}
\end{eqnarray}
and the independent variables are changed, via (\ref{U3}), from $(U,\zeta ,%
\overline{\zeta })$ to $(\phi ,\zeta ,\overline{\zeta }).$ They become

\begin{eqnarray}
\text{$\frak{d}$}_{|\phi }\text{(}V^{2}\phi _{1}^{0}) &=&0  \label{MaxI} \\
\text{$\frak{d}$}_{|\phi }\text{(}V\phi _{2}^{0}) &=&\partial _{\phi }\phi
_{1}^{0}  \label{Max2}
\end{eqnarray}
with $\frak{d}_{|\phi }$ meaning $\frak{d}$\ but holding $\phi ,$ rather
than $U$, constant and 
\begin{equation}
V=\partial _{\phi }X=\partial _{\phi }\xi ^{a}(\phi )\cdot l_{a}(\zeta ,%
\overline{\zeta }).  \label{V}
\end{equation}

The first can be integrated as 
\begin{equation}
\phi _{1}^{0}=Q(\phi )V^{-2}  \label{phi10}
\end{equation}
so that the second becomes 
\begin{equation}
\text{$\frak{d}$}_{|\phi }\text{(}V\phi _{2}^{0})=\partial _{\phi }\{Q(\phi
)V^{-2}\}.  \label{Max2*}
\end{equation}
When Eq.(\ref{Max2*}) is integrated over the sphere (at fixed $\phi $),
using the properties of $\frak{d}$, one has 
\[
\partial _{\phi }\{Q(\phi )\int V^{-2}d\Omega \}=0 
\]
but with $v^{a}=\partial _{\phi }\xi ^{a}$ chosen to be a norm 2 vector (by
the regauging of $\phi $) the integral term is constant and we have that

\[
Q(\phi )=\frac{q}{2}=\text{constant. } 
\]
(The factor 1/2 is chosen so that $q$ is the charge of a Coulomb field.)

The Eq.(\ref{Max2*}) can then be integrated, [using the $\phi $ derivative
of the identity, 
\begin{eqnarray*}
\text{V}^{2}(\text{$\frak{d}$}\overline{\text{$\frak{d}$}}\log V+1) &=&\frac{%
1}{2}\eta _{ab}\xi ^{\prime \,a}\xi ^{\prime \,b}=1 \\
v^{a} &\equiv &\xi ^{\prime \,a}\equiv \partial _{\phi }\xi ^{a}(\phi ),
\end{eqnarray*}
when the norm of $\xi ^{\prime \,a}$ is 2]\cite{NL,Lind}, as

\[
\phi _{2}^{0}=\frac{q}{2}V^{-1}\overline{\text{$\frak{d}$}}_{|\phi
}[V^{-1}\partial _{\phi }V]. 
\]

Since the condition for $l^{a}$ to be an eigenvector (pnv) of $F_{ab}$ is
that 
\begin{equation}
\phi _{0}=0,  \label{phi_0=0}
\end{equation}
our `shear-free' Maxwell field is then given by 
\begin{eqnarray}
\phi _{0} &=&0  \label{TheSolution} \\
\phi _{1} &=&\frac{q}{2}V^{-2}\rho ^{2}  \nonumber \\
\phi _{2} &=&\frac{q}{2}V^{-1}\overline{\text{$\frak{d}$}}_{|\phi
}[V^{-1}\partial _{\phi }V]\rho +O(\rho ^{2})  \nonumber
\end{eqnarray}
or, using the coordinate transformation, Eq.(\ref{nullgeodesics}), 
\[
F^{ab}(x^{a})=2\phi _{0}\overline{m}^{[a}n^{b]}-2\phi _{1}(l^{[a}n^{b]}+%
\overline{m}^{[a}m^{b]})+2\phi _{2}l^{[a}m^{b]}+c.c. 
\]
We see that the field is completely determined by the value of the charge $q$
and the complex function $\phi (U,\zeta ,\overline{\zeta }),$ via the
normalized complex-world-line by 
\begin{eqnarray}
z^{a} &=&\xi ^{a}(\phi )  \label{defines1} \\
U &=&\xi ^{a}(\phi )l_{a}(\zeta ,\overline{\zeta })  \label{defines2} \\
V &=&\partial _{\phi }\xi ^{a}(\phi )l_{a}(\zeta ,\overline{\zeta }).
\label{defines3}
\end{eqnarray}

The problem however is that it is virtually impossible to give a physical
meaning to this Maxwell field as it has been described above. In order to
give its physical meaning we would like to find the multipole moments by
means of a far-field analysis. To do this we must transform it into a more
useful coordinate/tetrad system and look at its asymptotic behavior. This
transformation need only be given asymptotically.

An associated issue is \textit{what is the real source of this Maxwell
field? }One sees immediately, from 
\[
\rho =-\frac{1}{r+i\Sigma }, 
\]
that the source has support only on the caustics of the congruence. In other
words, for the \textit{twisting, regular} null geodesic congruence, the
source is a closed string evolving in time. There is a subtle caveat
concerning the support of the source that we will discuss in the conclusion.

\subsection{The Asymptotic Field}

Our first task is to connect the coordinate and tetrad system associated
with the shear-free congruence with that of a Bondi system. If we start with
Minkowski space and its natural coordinates $x^{a}$ \thinspace the
transformation to Bondi coordinates ($U_{B},r_{B},\zeta _{B},\overline{\zeta 
}_{B}$) is given by 
\begin{eqnarray}
x^{a} &=&U_{B}t^{a}+r_{B}l_{B}^{a}  \label{BondiCoordinates} \\
l_{B}^{a} &=&\frac{\sqrt{2}}{2}(1,\frac{\zeta _{B}+\overline{\zeta }_{B}}{
1+\zeta _{B}\overline{\zeta }_{B}},-i\frac{\zeta _{B}-\overline{\zeta }_{B}}{
1+\zeta _{B}\overline{\zeta }_{B}},\frac{-1+\zeta _{B}\overline{\zeta }_{B}}{%
1+\zeta _{B}\overline{\zeta }_{B}}).  \nonumber
\end{eqnarray}

The other members of the tetrad are given by Eq.(\ref{tetrad}) with $\zeta $'%
$s$ replaced by the $\zeta _{B},$ etc. Note that these Bondi coordinates are
the special case, $L=0$, of the shear-free coordinates ($U,r,\zeta ,%
\overline{\zeta })$ of Eq.(\ref{null geodesics}). The transformation between
the two sets is given by equating, for the two transformations, their
respective $x^{a}$'$s;$%
\begin{equation}
U_{B}t^{a}+r_{B}l_{B}^{a}=Ut^{a}-\overline{L}m^{a}-L\overline{m}
^{a}+(r-r_{0})l^{a},  \label{toBondi}
\end{equation}

By transvecting with, respectively, $(t^{a},l^{a},m^{a},\overline{m}^{a}$),
and from the orthogonality properties, we have

\begin{eqnarray*}
\text{ }2U_{B}+r_{B} &=&2U+(r-r_{0}) \\
\text{ }U_{B}+r_{B}l_{B}\cdot l &=&U \\
\text{ }r_{B}l_{B}\cdot m &=&L \\
\text{ }r_{B}l_{B}\cdot \overline{m} &=&\overline{L}.
\end{eqnarray*}
Using the explicit scalar products

\begin{eqnarray*}
l\cdot l_{B} &\equiv &\frac{1}{(1+\zeta \overline{\zeta })(1+\zeta _{B}%
\overline{\zeta }_{B})}(\zeta -\zeta _{B})(\overline{\zeta }-\overline{\zeta 
}_{B}) \\
m\cdot l_{B} &\equiv &\frac{1}{(1+\zeta \overline{\zeta })(1+\zeta _{B}%
\overline{\zeta }_{B})}(\overline{\zeta }-\overline{\zeta }^{*})(1+\zeta _{B}%
\overline{\zeta }) \\
\overline{m}\cdot l_{B} &\equiv &\frac{1}{(1+\zeta \overline{\zeta }
)(1+\zeta _{B}\overline{\zeta }_{B})}(\zeta -\zeta _{B})(1+\overline{\zeta }
_{B}\zeta ) \\
t\cdot l_{B} &=&t\cdot l=1,
\end{eqnarray*}
we have the four coordinate transformations, given implicitly by 
\begin{eqnarray*}
\text{ }2U_{B}+r_{B} &=&2U+(r-r_{0}) \\
\text{ }U_{B}+\frac{L\overline{L}}{r_{B}} &=&U \\
\frac{1}{(1+\zeta \overline{\zeta })(1+\zeta _{B}\overline{\zeta }_{B})}(%
\overline{\zeta }-\overline{\zeta }_{B})(1+\zeta _{B}\overline{\zeta }) &=&%
\frac{L}{r_{B}} \\
\frac{1}{(1+\zeta \overline{\zeta })(1+\zeta _{B}\overline{\zeta }_{B})}
(\zeta -\zeta _{B})(1+\overline{\zeta }_{B}\zeta ) &=&\frac{\overline{L}}{
r_{B}}
\end{eqnarray*}
and their explicit asymptotic form

\begin{eqnarray}
\zeta _{B} &=&\zeta -\frac{\overline{L}}{r}(1+\zeta \overline{\zeta }
)+O(r^{-2})  \label{asypBondi} \\
\overline{\zeta }_{B} &=&\overline{\zeta }-\frac{L}{r}(1+\zeta \overline{
\zeta })+O(r^{-2})  \nonumber \\
r_{B} &=&(r-r_{0})+2\frac{L\overline{L}}{r}+O(r^{-2})  \nonumber \\
U_{B} &=&U-\frac{L\overline{L}}{r}+O(r^{-2}).  \nonumber
\end{eqnarray}

By returning to Eq.(\ref{toBondi}), using Eq.(\ref{asypBondi}), and 
\begin{equation}
t^{a}=l^{a}+n^{a}=l_{B}^{a}+n_{B}^{a}  \label{t}
\end{equation}
we have the first leg of the tetrad transformation 
\begin{equation}
\text{ }l_{B}^{a}(\zeta _{B},\overline{\zeta }_{B})=l^{a}-\frac{\overline{L}%
}{r}m^{a}-\frac{L}{r}\overline{m}^{a}+O(r^{-2}).  \label{l*}
\end{equation}
From Eq.(\ref{t}), we obtain the second leg, 
\begin{equation}
n_{B}^{a}=n^{a}+\frac{\overline{L}}{r}m^{a}+\frac{L}{r}\overline{m}%
^{a}+O(r^{-2}).  \label{n*}
\end{equation}
The $m^{*a}$ and $\overline{m}^{*a}$ equations can be found from the
orthogonality relations, yielding

\begin{eqnarray}
m_{B}^{a} &=&m^{a}+r^{-1}L(l^{a}-n^{a})+O(r^{-2})  \label{m*} \\
\overline{m}_{B}^{a} &=&\overline{m}^{a}+r^{-1}\overline{L}
(l^{a}-n^{a})+O(r^{-2}).  \nonumber
\end{eqnarray}
As the calculation proceeded it appeared as if higher order terms in $r^{-1}$
would be needed. It however turns out that to find the dipole moments these
terms are sufficient.

Denoting the spin-coefficient version of the Maxwell field in a Bondi
coordinate/tetrad system by

\begin{eqnarray}
\phi _{B0} &=&F_{ab}l_{B}^{a}m_{B}^{b}  \label{bondiphis} \\
\phi _{B1} &=&\frac{1}{2}F_{ab}(l_{B}^{a}n_{B}^{b}+m_{B}^{a}\overline{m}
_{B}^{b})  \nonumber \\
\phi _{B2} &=&F_{ab}\overline{m}_{B}^{a}n_{B}^{b},  \nonumber
\end{eqnarray}
and using

\[
l_{B}\wedge m_{B}=l\wedge m-r^{-1}L[\overline{m}\wedge m+l\wedge
n]+r^{-2}L^{2}\overline{m}\wedge n+... 
\]
(plus terms of order $r^{-2}$ and higher that make no contribution to the
dipole) with analogous expressions for the other bivectors, we have that 
\begin{eqnarray}
\phi _{B0} &=&-2\frac{L}{r}\phi _{1}+\frac{L^{2}}{r^{2}}\phi _{2}+O(r^{-4})
\label{phiTophi*} \\
\phi _{B1} &=&\phi _{1}-\frac{L}{r}\phi _{2}+O(r^{-3})  \nonumber \\
\phi _{B2} &=&\phi _{2}+O(r^{-2})  \nonumber
\end{eqnarray}
At this point we have used $\phi _{0}=0$ and, from Eq.(\ref{TheSolution}),
the known asymptotic behavior of the $\phi _{1}$ and $\phi _{2}$ with

\[
\rho =-\frac{1}{r+i\Sigma }=-\frac{1}{r}(1-\frac{i\Sigma }{r}+...)=-\frac{1}{
r}+\frac{i\Sigma }{r^{2}}+.... 
\]

When the twisting solution, Eq.(\ref{TheSolution}), is inserted in Eq.(\ref
{phiTophi*}) they become

\begin{eqnarray}
\phi _{B0} &=&r^{-3}\phi _{B0}^{0}+O(r^{-4})  \label{phiB} \\
\phi _{B1} &=&r^{-2}\phi _{B1}^{0}+O(r^{-3})  \nonumber \\
\phi _{B2} &=&r^{-1}\phi _{B2}^{0}+O(r^{-2})  \nonumber
\end{eqnarray}
with 
\begin{eqnarray}
\phi _{B0}^{0} &=&-q(LV^{-2}+\frac{1}{2}L^{2}V^{-1}\overline{\text{$\frak{d}$%
}}_{|\phi }[V^{-1}V,_{\phi }])  \label{phi^0*} \\
\phi _{B1}^{0} &=&\frac{q}{2V^{2}}(1+LV\overline{\text{$\frak{d}$}}_{|\phi
}[V^{-1}V,_{\phi }])  \label{phi^1*} \\
\phi _{B2}^{0} &=&-\frac{q}{2}V^{-1}\overline{\text{$\frak{d}$}}_{|\phi
}[V^{-1}V,_{\phi }]  \label{phi^3*} \\
&=&-\frac{q}{2}V^{-2}\{\partial _{\phi \phi }\xi ^{a}(\phi )\cdot \overline{m%
}_{a}(\zeta ,\overline{\zeta })-V^{-1}\partial _{\phi \phi }\xi ^{a}(\phi
)l_{a}\partial _{\phi }\xi ^{a}(\phi )\cdot \overline{m}_{a}(\zeta ,%
\overline{\zeta })\}  \label{**}
\end{eqnarray}

with all the quantities obtained from

\begin{eqnarray}
U &=&\xi ^{a}(\phi )l_{a}(\zeta ,\overline{\zeta }).  \label{U^} \\
L(U,\zeta ,\overline{\zeta }) &=&\xi ^{a}(\phi )m_{a}(\zeta ,\overline{\zeta 
}),  \label{L^} \\
V(U,\zeta ,\overline{\zeta }) &=&\partial _{\phi }X=\partial _{\phi }\xi
^{a}(\phi )\cdot l_{a}(\zeta ,\overline{\zeta })  \label{V^} \\
V,_{\phi } &=&\partial _{\phi \phi }\xi ^{a}(\phi )\cdot l_{a}(\zeta ,%
\overline{\zeta }).  \label{A^}
\end{eqnarray}

There are several important things to notice about these expressions.

$1.$ Though they are functions of the shear-free congruence coordinates, ($%
U,\zeta ,\overline{\zeta }),$ since we are in the asymptotic limit, the
transformation between the two types becomes the identity, the ($U,\zeta ,%
\overline{\zeta })$ can all be replaced simply by the Bondi coordinates, ($%
U_{B},\zeta _{B},\overline{\zeta }_{B}).$ In the remaining discussion we
will take that for granted and simply use ($U,\zeta ,\overline{\zeta })$ as
the Bondi coordinates.

2. The fact that the coordinate $U$ enters the fields only implicitly, via
Eq.(\ref{U^}), creates considerable difficulty in the interpretation. In
principle Eq.(\ref{U^}) should be inverted as

\begin{equation}
\phi =\Phi (U,\zeta ,\overline{\zeta })  \label{inverse}
\end{equation}
and substituted into Eqs.(\ref{V^}), (\ref{L^}) and (\ref{A^}). It is
possible to explicitly do this only in the simplest cases.

3. Clues to the possible physical meaning of these solutions can be obtained
by the following observations:

[i] In the real L.W. case, we can, at any one instant of proper-time, $\phi
_{re}=\tau =\tau _{0},$\textit{\ put the spatial coordinate origin on the
center of mass world line and chose, by a boost, the rest-frame of the
world-line, i.e.,} $V=1$ and $\xi ^{i}=0.$ \textit{\ }At that instant, the
radiation term becomes 
\begin{equation}
\phi _{2}^{0*}=-\frac{q}{2}\overline{\text{$\frak{d}$}}_{|\tau }[V,_{\tau
}]=-\frac{q}{2}\partial _{\tau }^{2}\xi ^{a}(\tau _{0})\cdot \overline{m}%
_{a}(\zeta ,\overline{\zeta }),  \label{electDipole}
\end{equation}
which is the known expression for \textit{pure} electric dipole radiation
with the electric dipole moment `\textit{identified '} as the spatial parts
of $q\xi ^{a}.$

\begin{remark}
Note that if we did not put \textit{origin on the center of mass world line
and chose the particles rest-frame}, i.e., $V=1,$ there still would be
electric dipole radiation (determined by $\partial _{\tau }^{2}\xi ^{a}(\tau
)$) but\textit{\ it would not be pure}, i.e., other multiple moments would
appear in the radiation. We will return to this later.
\end{remark}

[ii] Again with $V=1$ and with $\tau $ as proper-time, in the case of zero
acceleration, i.e., with $\partial _{\tau }^{2}\xi ^{a}(\tau )=0,$ the
``dipole aspect'', $\phi _{B0}^{0}$ is

\[
\phi _{B0}^{0}=-qL=-q\xi ^{a}(\tau )m_{a}(\zeta ,\overline{\zeta }), 
\]
with the \textit{spatial} part of $q\xi ^{a}$ being the dipole moments in
agreement with the standard definition.

\section{Complex Dipole Moment}

In general, when electromagnetic radiation is present, finding, or even
defining, the dipole moments (electric and magnetic) is difficult and is, in
general, not an invariant concept. For an arbitrary given time-dependent
charge and current distribution, from the usual definitions, the electric
and magnetic dipole moments can be calculated by integrals over the sources
at any one instant of time, in any one Lorentz frame. But, in any other
Lorentz frame, the new pair of dipole moments would have little relationship
to the first set. Assuming that the total charge is different from zero, one
could try to eliminate the ambiguity by choosing the Lorentz frame in which
the center of charge world-line is instantaneously at rest and then doing
the dipole calculation. This however presents other ambiguities; finding the
center of charge world-line also has problems and ambiguities. Though there
is a limiting process that appears to converge to a unambiguous world-line
it remains a cumbersome process and probably has little value. The dipole
moments in any case would depend on the spatial coordinate origin. However
in the case of the L.W. field, the center of charge world-line, 
\[
x^{a}=\xi ^{a}(\tau ), 
\]
coincides with the particle world-line and is thus unambiguous. At any
instant of time one could go to the particle rest-frame (with, of course, an
ambiguity in the choice of spatial origin) and define the electric dipole
moment by 
\begin{eqnarray}
d_{e}^{a}(\tau _{0}) &=&q\xi _{\perp }^{a}(\tau _{0})  \label{dipole.elect}
\\
\xi _{\perp }^{a}(\tau _{0}) &=&\xi ^{a}(\tau _{0})-\frac{1}{2}\partial
_{\tau }\xi ^{a}(\xi _{b}\partial _{\tau }\xi ^{b}),  \nonumber
\end{eqnarray}
with (in the rest-frame) 
\[
\partial _{\tau }\xi ^{a}=v^{a}=\sqrt{2}\delta _{0}^{a}. 
\]
We will simply take as our definition (in any Lorentz frame) Eq.(\ref
{dipole.elect}) as the definition of the electric dipole moment.

An alternate means of defining the dipole moments is to, first solve the
Maxwell equations, then look at the asymptotic fields. Normally, in the
static or stationary case, the dipole moments are found in the coefficients
of the $l=1$ spherical harmonics, in the r$^{-3}$ terms in the Maxwell field
and their second time derivatives appear in the $l=1$ radiation terms. Our
definition, Eq.(\ref{dipole.elect}), coincides with these asymptotic
identifications.

Turning now to the complex L.W. fields, we can see the unity in \textit{%
defining }the complex dipole moment, at any complex proper time $\phi ,$ by 
\begin{eqnarray*}
d_{c}^{a}(\phi ) &=&d_{e}^{a}(\phi )+id_{m}^{a}(\phi )=q\xi _{\perp
}^{a}(\phi ) \\
\xi _{\perp }^{a}(\phi ) &=&\xi ^{a}(\phi )-\frac{1}{2}\partial _{\phi }\xi
^{a}(\xi _{b}\partial _{\phi }\xi ^{b}).
\end{eqnarray*}
The argument or defense of this is to note that the complex L.W. Maxwell
fields, Eqs.(\ref{phi^0*}), (\ref{phi^1*}) and (\ref{phi^3*}) treat the $\xi
^{a}(\phi )$ as a single entity. They becomes the real L.W. when $\xi
^{a}(\phi )=\xi _{R}^{a}(\phi ).$ The complex dipole moments coincide with
the asymptotic definitions of the electric and magnetic moments in the
stationary case. All the equations (e.g., the radiation terms and the dipole
aspect terms) remain exactly unchanged in formal appearance from the real
L.W. case, except the $\xi ^{a}(\tau )$ is replaced $\xi ^{a}(\phi ).$ The
generalization is thus quite natural.

To explicitly calculate the electric and magnetic dipole parts of the
radiation in terms of the real and imaginary parts of the complex
acceleration, one must first express the acceleration in terms of $U$ rather
than $\phi .$ From $U=\xi ^{a}(\phi )l_{a}(\zeta ,\overline{\zeta }),$ one
solves for $\phi =\Phi (U,\zeta ,\overline{\zeta }),$ then replaces the $%
\phi $ in $\phi _{B2}^{0},$ Eq.(\ref{**}), and finally integrates the
expression over the $(\zeta ,\overline{\zeta })$ sphere weighted by the $%
l=1,s=1,$ spherical harmonics. This procedure is exactly the same for both
the real and complex cases and, in general, yields the dipole radiation as a
complex quantity whose real and imaginary values are respectively the
electric and magnetic parts.

\section{The Source}

We have until this point been discussing only the vacuum region of the
complex or real L.W. field. The question then can be raised: what is the
source of the field? From the complex point of view the source is obviously
the complex line, $z^{a}=\xi ^{a}(\phi ).$ But of more interest to us is the
real source. By looking at the fields, Eqs.(\ref{phi=0}) and (\ref{sol.I})

\begin{eqnarray}
\phi _{0} &=&0 \\
\phi _{1} &=&\rho ^{2}\phi _{1}^{0} \\
\phi _{2} &=&\rho \phi _{2}^{0}+\rho ^{2}\phi _{2}^{1}+\rho ^{3}\phi _{2}^{2}
\nonumber
\end{eqnarray}
with 
\begin{equation}
\rho =-\frac{1}{r+i\Sigma }  \nonumber
\end{equation}
and knowing that all the coefficients are regular functions of ($U$,$\zeta ,%
\overline{\zeta }),$ we immediately see that the solutions are singular
where $\rho $ is singular, i.e., on the caustics of the congruence. [See Eq.(%
\ref{caustic}).]. The source is thus a charged current `string'. (The
special stationary case, the singular circle, has been analyzed in great
detail by G. Kaiser\cite{GK} and others\cite{others,WI1}.) There is however,
as mentioned earlier, a caveat. In order to make the solutions single valued
on the real Minkowski space (or, equivalently, to avoid a double sheeted
Minkowski space) one must put in a charge distribution on the 2-sheet
bounded by the `string'.

It must be emphasized that these sources were never put in by hand; they
arose simply as the \textit{real} singularities of our complex L.W. fields.

\section{Examples}

\subsection{The Kerr Congruence;}

Beginning with the complex world line 
\[
z^{a}=\xi ^{a}(\varphi )=\xi _{0}^{a}+\varphi (\frac{\sqrt{2}}{2}
,0,0,0)=(0,0,0,i\frac{\sqrt{2}}{2}a)+\varphi \frac{\sqrt{2}}{2}(1,0,0,0) 
\]
we have that

\begin{eqnarray}
U &=&\xi ^{a}l_{a}=\frac{1}{2}\{ia(\frac{1-\zeta \overline{\zeta }}{1+\zeta 
\overline{\zeta }})+\varphi \}  \label{Kerr1} \\
\varphi &=&2U-ia(\frac{1-\zeta \overline{\zeta }}{1+\zeta \overline{\zeta }})
\label{Kerr2} \\
L &=&-ia[\frac{\overline{\zeta }}{(1+\zeta \overline{\zeta })}]\text{ \& }%
\overline{L}=ia[\frac{\zeta }{(1+\zeta \overline{\zeta })}]  \label{Kerr3} \\
\Sigma &=&a[\frac{1-\zeta \overline{\zeta }}{(1+\zeta \overline{\zeta })}]%
\text{ \& }r_{0}=0  \label{Kerr4}
\end{eqnarray}

The caustic and singularity of the Maxwell field occurs on the circle in the
(x,y) plane, 
\[
(x^{1})^{2}+(x^{2})^{2}=a^{2}, 
\]
the well-known Kerr singularity.

\subsection{The Moving Kerr congruence;}

\begin{eqnarray*}
\xi ^{a} &=&\varphi \frac{\sqrt{2}}{2}(1,0,0,B) \\
2U &=&\varphi (\frac{B(1-\zeta \overline{\zeta )}+1+\zeta \overline{\zeta }}{%
1+\zeta \overline{\zeta }}) \\
\varphi &=&\frac{2U(1+\zeta \overline{\zeta })}{1+B+\zeta \overline{\zeta }%
(1-B)} \\
L &=&-B\varphi \frac{\overline{\zeta }}{1+\zeta \overline{\zeta }}=-\frac{2BU%
\overline{\zeta }}{1+\zeta \overline{\zeta }+B(1-\zeta \overline{\zeta })} \\
\text{$\frak{d}$}\overline{L}+L\overline{L}_{U} &=&-\frac{2\overline{B}U}{%
(1+\zeta \overline{\zeta }+\overline{B}(1-\zeta \overline{\zeta }))^{2}}[%
1-(\zeta \overline{\zeta })^{2}+\overline{B}(1+(\zeta \overline{\zeta })^{2})%
] \\
&&+\frac{4UB\overline{B}\overline{\zeta }\zeta }{1+\zeta \overline{\zeta }%
+B(1-\zeta \overline{\zeta })}\frac{1}{1+\zeta \overline{\zeta }+\overline{B}%
(1-\zeta \overline{\zeta })}
\end{eqnarray*}

The vanishing of the twist, $\Sigma ,$ (a fairly complicated expression in
terms of $\zeta $ and $B)$ occurs when 
\[
\zeta \overline{\zeta }=\sqrt{\frac{1+(B^{*}+B)+B^{*}B}{1-(B^{*}+B)+B^{*}B}}
. 
\]
Thus, at any one time, the caustic is topologically a circle, lying parallel
to the (x,y) plane whose radius is a linear function of $U.$ When $B=ib$ the
circle is in the (x,y) plane; when $B=a<1$ it degenerates to a point.

\subsection{The Displaced moving Kerr congruence:}

\begin{eqnarray*}
\xi ^{a} &=&(0,0,\frac{\sqrt{2}}{2}ib,0)+\varphi (1,0,0,i\frac{\sqrt{2}}{2}b)
\\
2U &=&\varphi (1+ib\frac{1-\zeta \overline{\zeta }}{1+\zeta \overline{\zeta }%
})-a\frac{\zeta -\overline{\zeta }}{1+\zeta \overline{\zeta }} \\
\varphi &=&\frac{2U(1+\zeta \overline{\zeta })+a(\zeta -\overline{\zeta })}{%
(1+\zeta \overline{\zeta }+ib(1-\zeta \overline{\zeta }))} \\
L &=&-\frac{1}{2}\frac{a(1+\overline{\zeta }^{2})}{(1+\zeta \overline{\zeta }%
)}-\frac{iab\overline{\zeta }(\zeta -\overline{\zeta })}{(1+\zeta \overline{
\zeta })[1+\zeta \overline{\zeta }+ib(1-\zeta \overline{\zeta })]}-\frac{2Uib%
\overline{\zeta }}{[1+\zeta \overline{\zeta }+ib(1-\zeta \overline{\zeta })]}%
.
\end{eqnarray*}

It did not appear to be worthwhile to study in detail the caustics of this
congruence. It is however clear from the fact that there are no
singularities in the angular behavior of $L$ (and hence also in $\Sigma $
and $r_{0}$) that the caustics will be confined to a finite spacial region -
presumably a distorted circle.

\subsection{The Robinson congruence:}

As a last example we discuss the important class of null geodesics known as
the Robinson congruences. This class can be constructed by choosing a linear
function in Eq.(\ref{twistor}) to determine $L(U,\zeta ,\overline{\zeta }),$%
i.e., from 
\[
F(u-L\zeta ,\text{ }L+u\overline{\zeta },\text{ }\overline{\zeta })=a+b%
\overline{\zeta }+c(L+u\overline{\zeta })+d(u-L\zeta )=0. 
\]

We will see that in general, though they are caustics-free, they do have a
non-obvious and non-standard singular behavior that creates an angular
singularity for the associated Maxwell field. We can begin with the complex
world-line depending on the four complex parameters, ($A,B,C,D$);

\begin{eqnarray*}
\xi ^{a} &=&(AL^{a}-BM^{a})+\varphi \{CL^{a}-D\overline{M}^{a}\} \\
L^{a} &=&\frac{\sqrt{2}}{2}(1,0,0,1),\text{ }M^{a}=\frac{\sqrt{2}}{2}
(0,1,-i,0),\text{ }\overline{M}^{a}=\frac{\sqrt{2}}{2}(0,1,i,0)
\end{eqnarray*}
or

\[
\xi ^{a}=\frac{\sqrt{2}}{2}\{(A,-B,iB,A)+\varphi (C,-D,-iD,C). 
\]

This leads immediately to

\[
U=\frac{1}{1+\zeta \overline{\zeta }}\{A+B\overline{\zeta }+\varphi
(C+D\zeta )\}. 
\]
If $D$ were zero, then by regauging, $\varphi ^{*}=A+B\overline{\zeta }
+\varphi C,$ we have that 
\begin{eqnarray*}
U &=&\frac{\varphi ^{*}}{1+\zeta \overline{\zeta }} \\
\text{i.e.},\text{ now with }\xi ^{a} &=&\varphi ^{*}L^{a}.
\end{eqnarray*}
In this `singular' case the congruence is the twist-free congruence based on
the light-cones from a real null geodesic world-line and is not interesting
from our point of view. The maxwell fiedl is a L.W. field from a charged
particle moving on the null geodesic.

If $D\neq 0,$ we can take (by regauging) $C=1$ and obtain the general
Robinson congruence; 
\begin{eqnarray}
U &=&\xi ^{a}l_{a}=\frac{1}{(1+\zeta \overline{\zeta })}\{(A+B\overline{%
\zeta })+\varphi (1+D\zeta )\}  \label{U*} \\
\varphi &=&\frac{U(1+\zeta \overline{\zeta })-(A+B\overline{\zeta })}{%
(1+D\zeta )}  \label{p**} \\
L &=&-\frac{\text{$\frak{d}$}\varphi }{\varphi _{u}}=\frac{u(D-\overline{%
\zeta })-D(A+B\overline{\zeta })}{(1+D\zeta )}  \label{L**} \\
2i\Sigma &=&[BD-\overline{D}\overline{B}+D\overline{D}(\overline{A}-A)]\frac{%
(1+\zeta \overline{\zeta })}{(1+D\zeta )(1+\overline{D}\overline{\zeta })}
\label{sig*} \\
r_{0} &=&\frac{U[(1-\overline{\zeta }\zeta )\{1-D\overline{D}\}+2(\overline{%
\zeta }\overline{D}+D\zeta )]}{(1+\overline{D}\overline{\zeta })(1+D\zeta )}
\label{r0**} \\
&&-\frac{(AD\zeta +\overline{D}\overline{A}\overline{\zeta })+(\overline{B}%
\zeta +B\overline{\zeta })D\overline{D}}{(1+\overline{D}\overline{\zeta }%
)(1+D\zeta )}-\frac{1}{2}\frac{(\overline{A}+A)D\overline{D}-BD-\overline{B}%
\overline{D}}{(1+\overline{D}\overline{\zeta })(1+D\zeta )}[1-\overline{%
\zeta }\zeta ]  \nonumber
\end{eqnarray}
One sees immediately, from Eq.(\ref{sig*}), that aside from a few degenerate
values of the constants, $\Sigma $ does not vanish and that, in general, the
Robinson congruence does not have caustics. It does however have a different
unusual type of pathology. In the direction, $\zeta =-D^{-1},$ the twist is
singular and the complex divergence $\rho \ $goes to zero$.$ In the same
direction, $r_{0}$ is also singular so that the coordinate transformation to
the congruence coordinates $(U,r,\zeta ,\overline{\zeta })$ breaks down and
we have an angular (wire) singularity in the associated Maxwell field.

In the future we plan to study congruences with non-vanishing accelerations

\section{Conclusions and Conjectured Generalizations to GR}

Our present discussion of Maxwell fields was based on our choice of a
Maxwell principal null vector field with the property of being the tangent
field of a diverging, shear-free null geodesic congruence. This property of
geodesic congruences has an important and honored place in the study of the
vacuum Einstein equations.

The idea of principal null vector (pnv) fields extends naturally to general
relativity, where they are defined from algebraic properties of the Weyl
tensor. In general, there are four independent pnv fields. There are however
degenerate situations where several of them might coincide. The class of
vacuum (or Einstein-Maxwell) metrics with coinciding pnv's, referred to as
the algebraically special metrics, breaks down into many subclasses. We will
be concerned with the subclass referred to in the Petrov-Pirani-Penrose
classification scheme as the type [2,1,1] (or the very restricted class
[2,2]). The ''2'' refers to the fact that there is a degenerate pair. From
the Goldberg-Sachs theorem\cite{G.S.}, one knows that the degenerate pnv
field is the tangent field to a shear-free, null geodesic congruence. If we
add the further requirement that it also have non-vanishing divergence, we
have the generalization to GR of our special class of Maxwell fields.

It has already been pointed out by several authors\cite{LW1,LW2} that the
diverging, twist-free, [2,1,1] vacuum metrics are quite analogous to the
real L.W., Maxwell fields. We conjecture that the twisting ones are
analogous with the complex L.W. fields, the spin-angular momentum being the
analogue of the magnetic moments. Further, with the Einstein-Maxwell
equations, we conjecture that if the degenerate [2,1,1] Weyl pnv coincides
with the Maxwell pnv the solution will possess both a spin-angular momentum
and a magnetic moment.

We already know\cite{NL,Lind,Tal} that this class of metrics, (vacuum or
Einstein-Maxwell) will be determined by two parameters, the mass (and
charge) plus the function, $L(U,\zeta ,\overline{\zeta }),$ that reduces to
our earlier flat-space version. In addition, though the differential
equations satisfied by $L$ are far more complicated than those discussed
earlier, Eq.(\ref{shearfree}), the solutions will \textit{contain} the same
four complex analytic functions $\xi ^{a}(\phi )$. We are thus led to the
conjectures;

Conjecture\textbf{\ I}: In the class of twisting [2,1,1] vacuum metrics, the
function $L(U,\zeta ,\overline{\zeta })$ determines the spin-angular
momentum which is obtained from the imaginary part of the `complex
world-line', $\xi ^{a}(\varphi ).$

In the special case of the Kerr metric the conjecture is true.

Conjecture\textbf{\ II: }For the Einstein-Maxwell equations, the same
complex function $L(u,\zeta ,\overline{\zeta })$ determines both a magnetic
moment and a spin angular moment with the same complex world line, $\xi
^{a}(\varphi ).$

In the case of the charged Kerr metric and Maxwell field, the conjecture is
again true\cite{etal}.

Conjecture III: It appears very likely that the gyromagnetic ratio will, for
this case, be precisely the Dirac value.

This also is true in the case of the charged Kerr metric\cite{etal}.

As the equations are extremely complicated, considerable more work is
required to verify and fully understand these considerations.

Two final comments are:

1. We point out that, from a totally different point of view, [namely from
the study\cite{Pf}, in GR, of rotating charged mass distributions] the Dirac
value of the gyromagnetic ration appears to be almost arise naturally in
other situations and appears to be stable.

2. It has been pointed out to us by G. Kaiser, that the point of view we are
espousing here, in different versions with different techniques and
different results, has been investigated independently by V. Kassandrov\cite
{VK} and A. Burinskii\cite{AB,BB}.

\section{Acknowledgments}

We extend grateful thanks to both Roger Penrose and Andrzej Trautman for
insights and several valuable suggestions. We happily acknowledge support
from the National Science Foundation under grant \#PHY-0244513.

\section{Appendix A}

An alternate more intuitive expression for the radiation term 
\[
\phi _{2}^{0*}=-\frac{q}{2}V^{-1}\overline{\text{$\frak{d}$}}_{|\phi
}[V^{-1}V,_{\phi }] 
\]
can be found.

First, from 
\[
\partial _{U}\xi ^{a}(\phi )=\partial _{\phi }\xi ^{a}(\phi )\phi ,_{{\small %
U}}=\partial _{\phi }\xi ^{a}(\phi )V^{-1} 
\]
and 
\begin{eqnarray}
\partial _{U}^{2}\xi ^{a}(\phi ) &=&\partial _{U}[\partial _{\phi }\xi
^{a}(\phi )\phi ,_{{\small U}}]=\partial _{\phi \phi }\xi ^{a}(\phi
)V^{-2}-\partial _{\phi }\xi ^{a}(\phi )V^{-2}\partial _{U}V  \label{A} \\
&=&\partial _{\phi \phi }\xi ^{a}(\phi )V^{-2}-\partial _{\phi }\xi
^{a}(\phi )V^{-3}\partial _{\phi \phi }\xi ^{a}(\phi )l_{a}  \nonumber \\
\partial _{U}^{2}\xi ^{a}(\phi )\overline{m}_{a} &=&V^{-2}\partial _{\phi
\phi }\xi ^{a}(\phi )\overline{m}_{a}-V^{-3}\partial _{\phi }\xi ^{a}(\phi )%
\overline{m}_{a}\partial _{\phi \phi }\xi ^{a}(\phi )l_{a}  \label{B}
\end{eqnarray}
we have that

\begin{eqnarray}
\phi _{2}^{0*} &=&-\frac{q}{2}V^{-1}\overline{\text{$\frak{d}$}}_{|\phi
}[V^{-1}V,_{\phi }]  \label{phi2*} \\
&=&-\frac{q}{2}\{V^{-2}\overline{\text{$\frak{d}$}}_{|\phi }V,_{\phi
}-V^{-3}V,_{\phi }\overline{\text{$\frak{d}$}}_{|\phi }V\}  \nonumber \\
&=&-\frac{q}{2}V^{-2}\{\partial _{\phi \phi }\xi ^{a}(\phi )\cdot \overline{m%
}_{a}(\zeta ,\overline{\zeta })-V^{-1}\partial _{\phi \phi }\xi ^{a}(\phi
)l_{a}\partial _{\phi }\xi ^{a}(\phi )\cdot \overline{m}_{a}(\zeta ,%
\overline{\zeta })\}  \nonumber
\end{eqnarray}
becomes, from Eq.(\ref{B}),

\begin{equation}
\phi _{2}^{0*}=-\frac{q}{2}\partial _{U}^{2}\xi ^{a}(\phi )\overline{m}_{a}.
\label{ph2**}
\end{equation}

Aside from the fact that $\xi ^{a}(\phi )$ contains ($\zeta ,\overline{\zeta 
}$ ) terms, via $\phi =\Phi (U,\zeta ,\overline{\zeta }),$ this is the
correct expression for the dipole radiation from a moving charge.

\end{document}